\documentclass[twoside,fleqn]{article}

\usepackage{epsf}
\usepackage{espcrc2}
\usepackage{amstex}
\usepackage{amssymb}

\title{
       \begin{flushright}\normalsize
            LA-UR-99-5480
       \end{flushright}
	\vskip -0.4 cm
	Spectrum of Mesons and Baryons with $b$ Quarks \thanks{
	This work was done in collaboration with A. Ali Khan, T. Bhattacharya, 
	S. Collins, C. Davies, C. Morningstar, 
	J. Shigemitsu and J. Sloan, and supported by the DoE Grand Challenges award 
	at the ACL at Los Alamos.}
        }
\author{Rajan Gupta\address{MS B-285, 
		Los Alamos National Lab, Los Alamos, New Mexico 87545, USA}
       }
\begin{document}

\begin{abstract}
We present highlights of the spectrum of mesons and baryons calculated
using NRQCD for heavy quarks and tadpole improved clover action for the light
quarks. 
\end{abstract}

\maketitle

\section{HEAVY-LIGHT MESONS}

\begin{table}[thbp]
\begin{center}
\setlength{\tabcolsep}{0.15cm}
\begin{tabular}{|lc|l|l|}
\hline
\multicolumn{2}{|c|}{state $(n\;J^P)$}&
\multicolumn{1}{c|}{Lattice}&
\multicolumn{1}{c|}{Expt.}\\
\hline
\multicolumn{4}{|c|}{heavy-light mesons}\\
\hline
$B    $    & $1({}0^-)$ &  5296(04)($^{-2}_{+3}$)     &    5279      \\
$     $    & $2({}0^-)$ &  5895(116)($^{+20}_{-32}$)  &    5860(*)   \\
$B^*  $    & $1({}1^-)$ &  5319(02)($^{+0}_{-2}$)     &    5325(1)   \\
$B^*_0$    & $1({}0^+)$ &  5670(37)($^{+16}_{-24}$)   &              \\
$B_1$      & $1({}1^+)$ &  5726(38)($^{+20}_{-29}$)   &    5698(12)  \\
$B^*_2$    & $1({}2^+)$ &  5822(45)($^{+27}_{-35}$)   &    5779(*)\protect\cite{lep1} \\
\hline
\multicolumn{4}{|c|}{heavy-strange mesons}\\
\hline
&&&\\[-12pt]
$B_s  $    & $1({}0^-)$ &  5385(15)($^{-6}_{+7}$)($^{+20}_{-0}$)    &    5375(6)   \\[2pt]
$     $    & $2({}0^-)$ &  5935(57)($^{+27}_{-38}$)($^{+9}_{-0}$)   &              \\[2pt]
$B_s^*$    & $1({}1^-)$ &  5412(14)($^{-4}_{+2}$)($^{+20}_{-0}$)    &    5422(6)   \\[2pt]
$B^*_{s0}$ & $1({}0^+)$ &  5742(27)($^{+14}_{-20}$)($^{+15}_{-0}$)  &              \\[2pt]
$B_{s1}$   & $1({}1^+)$ &  5804(31)($^{+17}_{-26}$)($^{+16}_{-0}$)  &    5853(15)  \\[2pt]
$B^*_{s2}$ & $1({}2^+)$ &  5878(26)($^{+23}_{-33}$)($^{+11}_{-0}$)  &              \\[2pt]
\hline
\end{tabular}
\end{center}
\caption{Mass estimates in MeV for various meson states.  The $b$
quark mass is fixed using the spin-averaged $\overline B(1S)$.  The
first error of the lattice data is statistical (this bootstrap
estimate includes uncertainties due to extrapolations in quark
masses), the second represents the scale uncertainty due to varying
$a^{-1}$ between 1.8 and 2.0 GeV, and for the strange mesons we quote 
a third error associated with the uncertainty in fixing the strange quark
mass. Preliminary experimental values are denoted by asterisks.  The
experimental numbers quoted against the $1^+$ states correspond to
broad resonances that are possibly a mixture of the different $P$
states. The corresponding lattice estimates are an unresolved
combination of the $1^+$ and $1^{+\prime}$ states.}
\label{tab:meson_summary}
\end{table}

The results for heavy-light mesons are summarized in
Table~\ref{tab:meson_summary}.  The results presented here are based
on the same statistical sample, consisting of 102 quenched
configurations at $\beta = 6.0$ with lattice size $16^3\times 48$, as
used in our study of $f_B$ and $f_{B_s}$~\cite{fBpaper}.  The details
of the NRQCD action, the evolution equation for calculating the heavy
quark propagator, the method used for setting the lattice scale, and
the fixing of light and strange quark masses are given
in~\cite{fBpaper}.  The lattice scale, determined from $M_\rho$, is
$1.92(7)$ GeV, and we consider the range $1.8-2.0$ GeV to determine
the associated uncertainty. The bare $b$ quark mass used in the
heavy-light analyses is $a M_b^0 = 2.31(12)$. From this we estimate $
m_{b}^{\overline{MS}} (m_b^{\overline{MS}}) =
4.35(10)({}^{-3}_{+2})(10)$ GeV. For hadrons containing strange
quarks, the central values correspond to fixing $m_s$ using $M_K$, and
for errors we use the difference between it and using
$m_s(M_{K^*})$. Details of the analyses of the spectrum will appear
soon in~\cite{spectrum}.

Overall, our estimates are in rough
agreement with experimental data, the one exception being the
hyperfine splittings as discussed below. To understand the various
mass splittings, we use the following qualitative picture in which the
mass of a heavy-light hadron is considered to be a sum of:
\begin{itemize}
\item{}
the pole mass of the heavy quark ($M_h$) which is 
$\sim 1.5$ GeV for the $c$ quark and $\sim 5.0$ GeV for the $b$; 
\item{}
the constituent mass of the light quarks which is approximately $300$ MeV
for the $u,d$ and $450$ MeV for the $s$ quark as inferred from the octet and 
decuplet light baryons. 
\item{}
for orbitally and radially excited states, an excitation
energy of the light quark, which we expect  to be of the order of
$\Lambda_{QCD}$; 
\item{}
the $O(\Lambda_{QCD}^2/M_h)$ contributions are the kinetic energy
of the heavy quark and  the heavy-light hyperfine energy   
$E_{\sigma_H \cdot \sigma_l} \approx 46$ MeV, inferred from the  
experimental $B^\ast-B$ splitting; 
\item{}
and a residual binding energy $E_{be}$ encapsulating
the remaining interactions which we expect to be small
($O(\Lambda_{QCD}^3/M_h^2)$).  
\end{itemize}
To isolate individual terms and estimate their size and dependence on
the quark masses we construct different linear combinations of meson
and baryon masses.

The spin-averaged splitting $\overline{M}_{B_s} - \overline{M}_{B_d}$
should be dominated by the difference of the strange and light quark
masses. We find $90(9)({}^{+5}_{-3})({}^{+20}_{-0})$ MeV (the first
error includes statistical and extrapolation in quark masses, the
second is due to the scale uncertainty, and the third is the variation
if $m_s$ is set using $M_{K^\ast}$ instead of $M_K$) whereas the
experimental value is $96(6)$ MeV. The splitting shows no significant
dependence on the heavy quark mass, consistent with the expectation
that the change in the difference in the kinetic energy of the heavy
quark is small.

By a judicious combination of operators and sources for quark
propagators we are able to obtain a signal for P-wave states. The
estimates for ${}^3P_2$ and ${}^3P_0$ are unambiguous as the operators
used to probe these states do not mix with other states. Thus, their 
masses and the splittings 
$B_2^\ast-B_0^\ast = 155(32)({}^{+9}_{-13})$ MeV and 
$B_{s2}^\ast-B_{s0}^\ast = 136(23)({}^{+10}_{-13})({}^{+0}_{-4})$ MeV, 
are predictions since the $P$ states have not been resolved
experimentally.  We have used ${}^3P_1$ and ${}^1P_1$ operators in the
$LS$ coupling scheme to probe the physical $1^+$ and $1^{+\prime}$
states. In this case each operator has a non-zero overlap with the two
physical states; consequently to get the physical masses it is
essential to get a signal in both the direct and mixed
correlators. Since the data show no reliable signal in the mixed
correlators, we do not have predictions for the masses of the physical 
states. In Table~\ref{tab:meson_summary}, we quote the result from the
${}^3P_1$ correlator as a rough estimate for the $1^+$ state.

Radial and orbital splittings are expected to be dominated by the
difference in kinetic energies of the heavy and light quarks. Of these, 
the light quark contributes the most
$O(\Lambda_{QCD}^2/m_{constituent}) \sim O(\Lambda_{QCD})$. We find
$602(86)({}^{+25}_{-35})$ and $559(55)({}^{+31}_{-38})({}^{+0}_{-12})$
MeV for the $2\,{}^1S_0-1\,{}^1S_0$ splitting in the $B$ and $B_s$
systems respectively. The preliminary experimental value for the $B$
is $581$ MeV~\cite{lep1}. 
For the spin-averaged $1P-1S$ splitting we find 
$457(31)({}^{+24}_{-35})$ MeV for the $B$, and
$428(27)({}^{+27}_{-41})({}^{+0}_{-2})$ MeV for the $B_s$.

The results for the hyperfine splittings are not as encouraging. We
find $\Delta E(B^\ast_d-B_d) = 24(5)({}^{+2}_{-3})$ MeV and $\Delta
E(B^\ast_s-B_s) = 27(3)({}^{+2}_{-3})({}^{+1}_{-0})$ MeV, $i.e.$
roughly half the experimental values, $46$ and $47$ MeV respectively.
Further work is required to clarify whether this discrepancy is due to
the quenched approximation or due to an underestimate of the $\vec
\sigma \cdot \vec B$ term (the clover term) in the action.

\section{BARYONS}

Our results for baryons are summarized in Table~\ref{tab:baryexp}. 
The first splitting we comment on is 
$ M_{\Lambda_h} - \left(M_H + 3M_{H^\ast}\right)/4 $. In this 
there is no contribution from the heavy quark mass and hyperfine interaction
$E_{\sigma_H \cdot \sigma_l}$, so it should be dominated by mass of the extra light quark in $\Lambda$.
The experimental values, $311(10)$ and $310(2)$ MeV for the $b$ and $c$
systems respectively, support this. Our lattice 
estimates are $\Lambda_b - \overline{B} = 370(67)({}^{+14}_{-20})$ MeV and
$\Xi_b - \overline{B_s} = 392(50)({}^{+15}_{-0})$ MeV. Also, 
the data show little dependence on the heavy quark mass. 

\begin{table}[thb]
\setlength\tabcolsep{5pt}
\begin{center}
\begin{tabular}{|c|c|c|}
\noalign{\hrule}
\multicolumn{1}{|c}{baryon} &  
\multicolumn{1}{|c}{ expt.} &
\multicolumn{1}{|c|}{Our results} \\
\noalign{\hrule}
\multicolumn{3}{|c|}{$\Lambda$-like ($s_l$ = 0, j= 1/2)} \\
\noalign{\hrule}
$\Lambda_h$      $(udb)$ & $5.624(9)$                     & $5.679(71)({}^{+14}_{-19})$ \\
$\Xi_h$          $(lsb)$ &                                & $5.795(53)({}^{+9}_{-15})(^{+15}_{-0})$ \\
\noalign{\hrule}	   				     
\multicolumn{3}{|c|}{$\Sigma$-like ($s_l$ = 1, j= 1/2)} \\			     
\noalign{\hrule}	   				     
$\Sigma_h$       $(llb)$ & $5.797(8)$~\protect\cite{lep2} & $5.887(49)({}^{+25}_{-37})$ \\
$\Xi_h^\prime$   $(lsb)$ &                                & $5.968(39)({}^{+20}_{-32})(^{+24}_{-0})$ \\
$\Omega_h$       $(ssb)$ &                                & $6.048(33)({}^{+16}_{-26})(^{+34}_{-0})$ \\
\noalign{\hrule}	   				     
\multicolumn{3}{|c|}{$\Sigma$-like ($s_l$ = 1, j= 3/2)} \\			     
\noalign{\hrule}	   				     
$\Sigma_h^\ast$  $(llb)$ & $5.853(8)$~\protect\cite{lep2} & $5.909(47)({}^{+25}_{-39})$ \\
$\Xi_h^\ast$     $(lsb)$ &                                & $5.989(39)({}^{+22}_{-34})(^{+25}_{-0})$ \\
$\Omega_h^\ast$  $(ssb)$ &                                & $6.069(34)({}^{+18}_{-30})(^{+35}_{-0})$ \\
\noalign{\hrule}
\end{tabular}
\end{center}
 \par\vfil\penalty-5000\vfilneg
\caption{Summary of $b$ baryons masses in GeV
($l$ stands for a $u$ or $d$ quark).}
\label{tab:baryexp}
\end{table}

In our picture the hyperfine interaction between the light quarks
should dominate the splitting $(2\Sigma_h +
4\Sigma^\ast_h)/6-\Lambda_h$.  A calculation of the $\sigma_l \cdot
\sigma_l$ term, assuming a simple non-relativistic model, suggests that
this splitting should be $2/3$ of the Delta-Nucleon splitting which
is $293$ MeV.  The experimental results, $(2\Sigma_c +
4\Sigma^\ast_c)/6-\Lambda_c = 212$ MeV and $(2\Sigma_b +
4\Sigma^\ast_b)/6-\Lambda_b= 210$ MeV (preliminary) are consistent
with this and suggest negligible dependence on the heavy quark mass.
We find $221(71)({}^{+12}_{-16})$ MeV at $M_b$ and no significant
dependence on the heavy quark mass. The value decreases to
$186(51)({}^{+13}_{-17})({}^{+0}_{-10})$ MeV on replacing $d$ with
$s$, in qualitative agreement with the experimental results in the
charmed sector.

The $\Sigma_h^\ast-\Sigma_h$ splitting is expected to be proportional
to $(1/M_h)$ as it should depend only on the heavy-light hyperfine
interaction $E_{\sigma_h \cdot \sigma_l}$.  Using this heavy quark
scaling one expects a value $(B^\ast - B) (\Sigma^\ast_c-\Sigma_c) /
(D^\ast - D) \approx 46 * 66/140 \approx 22 $ MeV. We find
$19(7)({}^{+2}_{-3})$ MeV, whereas the preliminary experimental value
is $56(8)$ MeV~\cite{lep2}. However, the experimental identification
of the states is still under debate.

\section{HEAVY-HEAVY MESONS $--$ ONIA}

\begin{table}[thb]
\setlength\tabcolsep{5pt}
\vskip -0.2cm
\begin{center}
\begin{tabular}{|c|c|c|c|}
\noalign{\hrule}
\multicolumn{1}{|c} {} &
\multicolumn{1}{|c} {} &
\multicolumn{1}{|c} {} &
\multicolumn{1}{|c|} {} \\[-10pt]
\multicolumn{1}{|c} {} &  
\multicolumn{1}{|c} { $a^{-1}$ MeV} &
\multicolumn{1}{|c} {$M_b^0 a$}      &
\multicolumn{1}{|c|}{$M_b^{\overline MS}$  GeV} \\
\noalign{\hrule}
                                    &             &              &            \\[-10pt]
$2{}^3S_1 - 1{}^3S_1$               & 2313(99)    &  $1.76(7)$   &  4.07(24)  \\
$2{}^1S_0 - 1{}^1S_0$               & 2413(83)    &  $1.69(5)$   &  4.08(19)  \\
${\overline {{}^3 P}}_1 - {}^3S_1$  & 2424(133)   &  $1.69(8)$   &  4.10(30)  \\
\noalign{\hrule}
\end{tabular}
\end{center}
 \par\vfil\penalty-5000\vfilneg
\caption{Estimates of lattice scale and $b$ quark mass from $\Upsilon$. The 
final column gives $M_b^{\overline MS}(M_b^{\overline MS})$.}
\vskip -0.2cm
\label{tab:upsilon}
\end{table}

The bottomonia spectrum has been successfully used for an independent
determination of the lattice scale and $m_b$~\cite{upsilon}. Here we 
present preliminary results based on the same data set as for the 
heavy-light analyses presented above.  
To extract the lattice scale we assume that the splittings
$(M_{\Upsilon^\prime} - M_{\Upsilon})$, 
$(M_{\eta_b^\prime} - M_{\eta_b})$, and 
$(M_{\overline{\chi}_b^\prime} - M_{\Upsilon})$, are independent 
of the heavy quark mass. We include $(M_{\eta_b^\prime} - M_{\eta_b})$ as 
it has the best signal and we assume it is equal to 
$(M_{\Upsilon^\prime} - M_{\Upsilon})$ on basis of the approximate 
equality in the charm system. Having fixed $1/a$ we determine $M_b^0$ by 
fixing $M_\Upsilon$ using the dispersion relation. These results are 
summarized in table~\ref{tab:upsilon}. 

Typically, the results for mass-splittings with best control over
statistical errors are obtained from fits to ratios of correlators. In
our data we find marginally better signal from fits to individual as
compared to ratio of correlators (a consequence of chosing a smearing
in heavy quark propagator generation that optimizes signal for
heavy-light states). For the hyperfine splitting $M_{\Upsilon} -
M_{\eta_b}$ we find $34.2(3.6)$ MeV (ratio fits) and $30.4(3.2)$ MeV
(difference of mass fits). If one scales the $J/\psi - \eta_c$
splitting, $127$ MeV, by $M_{J/\psi}/M_\Upsilon$ then one expects
$\sim 42$ MeV. Thus, once again the hyperfine splitting is
underestimated though not by as much as for $M_{B^\ast} - M_B$.

The determination of $P$ state splittings has larger errors. We find 
${}^1P_1 - {}^3 {\overline P} = 0(8)$ where ${}^3 {\overline P} = 
({}^3P_0 + 3*{}^3P_1 + 3*{}^3P_2 (T) + 2*{}^3P_2 (E) )/9$;  
${}^3P_0 - {}^3 {\overline P} = -42(23)$; 
${}^3P_1 - {}^3 {\overline P} = 13(23)$; 
${}^3P_2 (T) - {}^3 {\overline P} = 28(14)$; and 
${}^3P_0 (E) - {}^3 {\overline P} = -40(23)$ MeV. 
It is disturbing to find the $T$ and $E$ cubic representations of 
${}^3P_2$ give such different estimates. 
The corresponding experimental values are 
$-40.2$,  $-8.3$, and $13$ MeV. 

%

%\section*{Acknowledgements} 
%This research was supported by DOE grants ???.

%
% Okay, now the list of journals

\def\ie{{\sl i.e.}}
\def\etal{{\it et al.}}
\def\etc{{\it etc.}}
\def\ibid{{\it ibid}}

%%%%%%%%%%%%%%%%%%%%%%%%%%%%%%%%%%%%%%%%%%%%%%%%%%%%%%%%%%%%%%%%%%

\end{document}